\documentclass[10pt,letterpaper,twocolumn,english,aps,prb,showpacs]{revtex4-1}
\usepackage{float}
\usepackage{amsmath}
\usepackage{amssymb}
\usepackage{graphicx}

\begin{document}

\title{Impurity scattering in the bulk of topological insulators}

\author{Cheung Chan}

\author{Tai-Kai Ng}

\affiliation{Department of Physics, The Hong Kong University of Science and Technology,
Clear Water Bay, Hong Kong}
\begin{abstract}
We study in this paper time-reversal $\delta$-impurity scattering
effects in the bulk of topological insulators (TI) in two and three
dimensions. Specifically we consider how impurity scattering strength
is affected by the bulk band structure of topological insulators.
An interesting band inversion effect associated with the change of
the system from ordinary to topological insulator is pointed out.
Experimental consequences of our findings are discussed. 
\end{abstract}


\maketitle

\subsection{Introduction}

Topological insulators (TI) are insulators characterized by a $\mathbb{Z}_{2}$
topological number of the bulk band structure, and realize a new topological
quantum state protected by time-reversal symmetry\cite{RevModPhys.82.3045,RevModPhys.83.1057}.
Theoretically, TI are insulating in the bulk but possess gapless boundary
states of helical Dirac fermions that give rise to interesting transport
properties not realized in ordinary insulators (OI). Novel excitations
like Majorana zero modes\cite{PhysRevB.82.115120} have been proposed
to exist in this class of materials and applications of the materials
in spintronics have also been discussed\cite{Moore}.

Towards the applications of TI, a better understanding of the overall
properties of the materials beyond their topological surface excitations
is needed. For example, to utilize exotic boundary states of TI one
needs to ensure a sufficient insulating bulk resistivity, which is
a challenge at present due to the presence of bulk conductivity \cite{Ong-2009,Qu13082010,Analytis2010,Analytis-PRB2010,Ando-2010,Butch-2010}
largely arising from the presence of impurities and defects in the
materials\cite{PhysRevB.82.241306}. If the defects are external,
which is the case for doped materials, one way to attain high bulk
resistivity is to synthesize the materials with delicate balance between
donors and acceptors \cite{PhysRevB.84.165311,Ong-arxiv2011}. On
the other hand, we also want to understand how defects can affect
the bulk conductivity through impurity induced in-gap bound states
that lead to impurity band and effective narrowing of band gap.

It was pointed out in Ref.\ \onlinecite{PhysRevB.82.115120} that
point-like defects in TI do not give rise to topological protected
zero-energy states. However, in-gap bound states can still be induced
by (isolated) impurities\cite{PhysRevB.84.035307,2010arXiv1009.5502S,1367-2630-13-10-103016}
through conventional mechanism. For example, depletion of wavefunction
at the defect results in-gap bound states if one imagines that the
impurity is formed by bending and shrinking the edge into a localized
defect \cite{PhysRevB.84.035307}. However, unlike the (real) edge
states, these localized bound states are finite energy modes because
they suffer from symmetry allowed (self) interactions. In this paper
we consider isolated time-reversal impurities described by $\delta$-function
potentials, and study how impurity scattering can depend on the special
band structure of TI's. We shall show that impurity scattering can
be enhanced in the bulk TI compared with OI due to particular band
structure associated with the materials. A criteria to search for
materials with suppressed impurity scattering effect is given. An
interesting sign-inversion effect in impurity scattering associated
with the band inversion in TI-OI transition is also pointed out. Experimental
implications of these effects are discussed.

\subsection{Formulation}

We adopt the modified Dirac Hamiltonian in either two or three dimensions
(2D/3D) as effective models for the topological insulator\cite{PhysRevB.76.045302,2010arXiv1009.5502S,PhysRevB.84.035307,1367-2630-13-10-103016},
i,e, 
\begin{equation}
H=\sum_{\mathbf{k},s}\psi_{\mathbf{k}s}^{\dagger}h_{2(3)D}(\mathbf{k},s)\psi_{\mathbf{k}s}
\end{equation}
 where $\psi_{\mathbf{k}s}^{\dagger}(\psi_{\mathbf{k}s})$ are 2-component
Dirac fields with momentum and spin indices $\mathbf{k}$ and $s=\uparrow,\downarrow$,
respectively and 
\begin{equation}
h_{2D}(\mathbf{k},s)=k_{x}\sigma^{y}+k_{y}\sigma^{x}s^{z}+m_{k}\sigma^{z},\label{eq:diraceff}
\end{equation}
 for 2D topological insulators where $\sigma$'s and $s$'s are Pauli
matrices, and $m_{k}=m-Bk^{2}$ . $\sigma^{z}=+1(-1)$ describes two
contributing atomic orbitals to the topological insulator. For example,
they represent $s$ and $p$ states for HgTe/CdTe system. The model
can also be generalized to describe 3D topological insulators with
a similar effective Hamiltonian $h_{3D}(\mathbf{k},s)$ with $m_{k}=m-\sum_{i}B_{i}k_{i}^{2}$
where $i=x,y,z$ and $B_{i}$'s are parameters of the same sign. We
shall for simplicity consider $B_{x}=B_{y}=B\neq B_{z}$, corresponding
to a common type of topological insulators in 3D with crystal structure
$R3m$.

For $mB_{i}<0$, the bands are ordered conventionally throughout the
Brillouin zone and the system is a ordinary (topologically trivial)
insulator. The bands near $\mathbf{k}=0$ are inverted due to strong
spin-orbit coupling for $mB_{i}>0$ and the system becomes a topological
insulator. The energy eigenvalues of the bulk Hamiltonians are given
by $\omega_{k}=\pm\sqrt{k^{2}+\left(m-\sum B_{i}k_{i}^{2}\right)^{2}}$
and are doubly degenerate (Kramers degeneracy). The time reversal
operator is given by $\hat{\mathcal{T}}=is^{y}\hat{K}$ with $\hat{\mathcal{T}}^{2}=-1$,
where $\hat{K}$ is the complex conjugate operator.

In real materials there are other electronic bands and impurity scatterings
exist in general between the Dirac bands and the other bands. To capture
these effects we consider also inter-band scattering between the Dirac
bands with a model quadratic band described by 
\begin{equation}
H_{\text{q}}=\sum_{\mathbf{k}s}\left(\frac{k^{2}}{2M}+\mu\right)c_{\mathbf{k}s}^{\dagger}c_{\mathbf{k}s},
\end{equation}
 where $M$ and $\mu$ are chosen to have the same sign so that the
quadratic band does not cross the Fermi level and the system remains
a bulk insulator. $M,\mu>0$ ($<0$) refers to a conduction (valence)
band in this notation.

We shall describe the impurity scattering by a single $s$-wave delta
function potential sitting at the origin in this paper. In the four-band
modified Dirac model we adopt here, the impurity scattering term within
the topological bands can be written in orbital basis\cite{PhysRevLett.97.226801}
as 
\begin{equation}
H_{I1}=\hat{u}\delta(\vec{x}),
\end{equation}
 where $\hat{u}$ is a $4\times4$ matrix. For time reversal impurities
$\hat{u}$ should be Hermitian and time reversal symmetric, i.e. $\hat{\mathcal{T}}\hat{u}\hat{\mathcal{T}}^{-1}=\hat{u}$.
There are six bases for $\hat{u}$, denoted as $u^{(i=0,..,5)}$=
($I$, $\sigma^{x},$ $\sigma^{z}$, $\sigma^{y}s^{x}$, $\sigma^{y}s^{y}$,
$\sigma^{y}s^{z}$).

To describe scattering between the topological and quadratic bands,
we consider 
\begin{equation}
H_{I2}=\sum_{\sigma,s}v_{\sigma}\left(\Psi_{s}^{\dagger}(0)\psi_{\sigma s}(0)+\psi_{\sigma s}(0)^{\dagger}\Psi_{s}(0)\right),\label{eq:inter}
\end{equation}
 where $\sigma=\pm1$ and $s=\uparrow,\downarrow$. $\Psi_{s}(x)$
is the Fourier transform of $c_{\mathbf{k}s}$. We include only spin-independent
scattering in writing down $H_{I2}$.

The effect of impurity scattering is described by the $T$-matrix
defined by 
\begin{subequations}
\begin{equation}
\hat{T}(\omega)=\left(\mathbb{I}-\hat{U}\hat{G}_{0}(\omega)\right)^{-1}\hat{U},\label{t-matrix}
\end{equation}
 for the $s$-wave delta function potential, where $\hat{U}$ is the
impurity scattering matrix and $\hat{G}_{0}(\omega)=\sum_{\mathbf{k}}\hat{G}_{0}(\mathbf{k},\omega)$
is the corresponding on-site matrix Green's function. In particular,
the existence of impurity bound state is determined by the eigenvalue
equation\cite{RevModPhys.78.373,PhysRevB.80.104504} 
\begin{equation}
\det\left[\mathbb{I}-\hat{G}_{0}(\omega)\hat{U}\right]=0.\label{eq:BS_eq}
\end{equation}
 
\end{subequations}
For the modified Dirac model, the on-site matrix Green's function
in the basis $\left((1,1),(-1,1),(1,-1),(-1,-1)\right)$ (w.r.t.\,$(\sigma^{z},s^{z}))$
is given by 
\begin{equation}
\hat{G}_{0}(\omega)=\left[\begin{array}{cccc}
g^{+}(\omega)\\
 & g^{-}(\omega)\\
 &  & g^{+}(\omega)\\
 &  &  & g^{-}(\omega)
\end{array}\right],\label{eq:G_0}
\end{equation}
 where 
\begin{eqnarray}
g^{\pm}(\omega) & = & -\int d\Omega_{D}\int_{0}^{\Lambda}k^{D-1}dk\frac{\omega\pm m_{k}}{k^{2}+m_{k}^{2}-\omega^{2}},\label{eq:g^pm}
\end{eqnarray}
 where $\Omega_{D}$ is the solid angle in dimension $D=2,3$. $\Lambda$
is a high energy cutoff above which the effective Dirac model (\ref{eq:diraceff})
is inadequate in describing the band-structure. Notice that the on-site
Green's function matrix $\hat{G}_{0}$ involved in Eq.\ (\ref{eq:BS_eq})
is diagonal. This is a special feature of the delta-function impurity
potential and would be absent if the impurity potential includes higher
angular momentum components.

\subsection{$g^{\pm}(\omega)$ around band-edge and enhanced impurity scattering}

Without loss of generality, we shall assume $B>0$ in the following.
We note that in the limit of weak impurity scattering, impurity-induced
in-gap bound states can exist only if $g^{\pm}(\omega)$ diverges
near the band edge. In usual (2D) semi-conductors the divergence in
$g^{\pm}(\omega)$ is originated from the band extremum associated
with a quadratic band structure. In TI's, the band structure is more
complicated and the band extremum may locate at a line of $\vec{k}$
values leading to stronger divergence behavior in $g^{\pm}(\omega)$.
This phenomenon may arise in both 2D and 3D TI's and will lead to
enhanced impurity scattering compared with usual semiconductors. First
we consider 2D.

\subsubsection{2mB$<$1}

The integral \eqref{eq:g^pm} for $g^{\pm}(\omega)$ can be evaluated
exactly in 2D. For $2mB<1$ the band minimum is located at $k=0$
with band gap $|m|$ as in usual semi-conductors. It is straightforward
to show that in this case 
\begin{equation}
g^{\pm}(\omega)|_{|\omega|\rightarrow|m|^{-}}\rightarrow-\pi\left(\omega\pm m\right)\ln\left(1+\frac{\delta^{2}}{m^{2}-\omega^{2}}\right),\label{g2d1}
\end{equation}
 where $\delta\sim\min(\Lambda,\sqrt{\left|\frac{m}{B}\right|})$
is the momentum cutoff below which we may approximate $m_{k}\sim m$.
We note that a logarithmic divergence appears in $g^{\pm}(\omega)$
near the band edge when $|\omega|\rightarrow m^{-}$. Similar logarithmic
divergence occurs for quadratic bands with Green's function given
by 
\begin{equation}
g_{0}^{\text{q}}(\omega)=-2\pi M\ln\left(1+\frac{\Lambda^{2}}{2M(\mu-\omega)}\right).
\end{equation}
 The logarithmic divergence arises from the quadratic dispersion of
electron dispersion near the band edge and is responsible for appearance
of impurity bound state for arbitrarily weak attractive scattering
potential in 2D. Taking $B>0$, we find that $g^{+(-)}(\omega)$ has
logarithmic divergence at $\omega\rightarrow+(-)|m|$ to $-(+)\infty$
for topological insulators (TI) ($m>0$) and has logarithmic divergence
at $\omega\rightarrow-(+)|m|$ to $+(-)\infty$ for ordinary insulators
(OI) ($m<0$), implying that the role of $g^{+(-)}(\omega)$ is inverted
when the system changes from OI to TI. Microscopically, a system changes
from OI to TI when there is an orbital inversion ($2m=$ difference
in energy between the two orbitals forming the inverted bands) and
the inversion in the role of $g^{\pm}(\omega)$ is a direct consequence
of this orbital inversion. In other words, there is a direct correspondence
between $g_{\text{TI}}^{\pm}(\omega)\leftrightarrow g_{\text{OI}}^{\mp}(\omega)$
with the same band gap $|m|$.

\subsubsection{2mB$>$1}

In this case, the band extremum occurs at a line of momentum $\vec{k}$
with $|\vec{k}|=k_{0}$. The corresponding bandgap is given by $\Delta=\frac{\sqrt{4mB-1}}{2|B|}$.
Evaluating the integral, we find 
\begin{eqnarray}
g^{\pm}(\omega)|_{|\omega|\rightarrow\Delta^{-}} & \rightarrow & -\frac{\pi^{2}}{\sqrt{2|B|}}\frac{\omega\pm\frac{1}{2B}}{\sqrt{\Delta^{2}-\omega^{2}}}+\text{const.}\label{g2d2}
\end{eqnarray}
 We see that $g^{\pm}(\omega)$ has a stronger (inverse square-root)
divergence at the gap edge compared with the case $2mB<1$. This is
a result of {}``dimension reduction'' in the integral when the band-minimum
occurs at a ring of momentum $\vec{k}$ with $|\vec{k}|=k_{0}$. We
notice that $g^{+}(\omega)$ diverges to $-(+)\infty$ at $\omega\rightarrow+(-)\Delta$,
with a weaker divergence (weighting factor) at $\omega\rightarrow-\Delta$.
A similar situation occurs for $g^{-}(\omega)$ which diverges to
$+(-)\infty$ at $\omega\rightarrow-(+)\Delta$, with a weaker divergence
at $\omega\rightarrow\Delta$. The appearance of divergences at both
$\omega=\pm\Delta$ is possible because of hybridization between the
two atomic orbitals. We note that similar double-divergent behavior
cannot occur for ordinary insulators which exist only at $mB<0$.

\subsubsection{3D case}

\begin{figure}
\includegraphics[width=0.55\columnwidth]{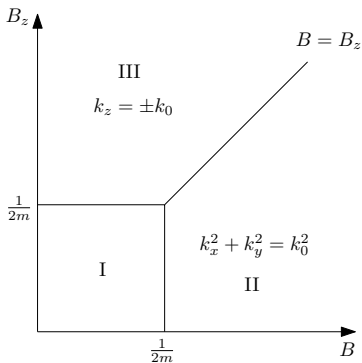}

\caption{Parameter range for band extrema $k_{0}$.\label{fig:band-min}}
\end{figure}

We can study $g^{\pm}(\omega)$ for 3D TI's in a similar way. Using
the Hamiltonian $h_{3D}(\vec{k},s)$ with $B_{x}=B_{y}=B\neq B_{z}$
we find that the divergent behavior of $g^{\pm}(\omega)$ can be classified
in three different regions as shown in Fig.\ \ref{fig:band-min}.
Region I is defined by $2m(B_{z},B)<1$ with band extremum occurring
at $\vec{k}=0$. Region II is defined by $B>\max\left(\frac{1}{2m},B_{z}\right)$
with band extremum occurring at a ring of $\vec{k}$ values, $\vec{k}=(k_{x},k_{y},0)$
with $k_{x}^{2}+k_{y}^{2}=k_{0}^{2}$ where $k_{0}=\frac{1}{B}\sqrt{mB-\frac{1}{2}}$.
Region III is defined by $B_{z}>\max\left(\frac{1}{2m},B\right)$
with band extrema occurring at $\vec{k}=(0,0,\pm k_{0})$ where $k_{0}=\frac{1}{B_{z}}\sqrt{mB_{z}-\frac{1}{2}}$.

The corresponding Green's functions are given by 
\begin{subequations}
\begin{align}
\frac{g_{\text{I}}^{\pm}(\omega)}{-4\pi} & \sim\int_{0}^{\delta}k^{2}dk\frac{\omega\pm m}{k^{2}+m^{2}-\omega^{2}}\nonumber \\
 & =(\omega\pm m)\left[\delta-\sqrt{m^{2}-\omega^{2}}\tan^{-1}\left(\frac{\delta}{\sqrt{m^{2}-\omega^{2}}}\right)\right]\nonumber \\
 & \sim\text{const. as }|\omega|\rightarrow|m|^{-}
\end{align}
 
\begin{eqnarray}
\frac{g_{\text{II}}^{\pm}(\omega)}{-2\pi} & \sim & \int_{k_{0}-\frac{\delta}{2}}^{k_{0}+\frac{\delta}{2}}dk\int_{0}^{\delta_{z}}dk_{z}\frac{k_{0}\left(\omega\pm\left(m-Bk_{0}^{2}\right)\right)}{k^{2}+k_{z}^{2}+m_{k}^{2}-\omega^{2}}\\
 & \sim & \frac{\omega\pm\frac{1}{2B}}{\sqrt{B(B-B_{z})}}\log\left(\frac{1}{\sqrt{\Delta^{2}-\omega^{2}}}\right)\text{ as }|\omega|\rightarrow\Delta^{-}\nonumber 
\end{eqnarray}
 
\begin{eqnarray}
\frac{g_{\text{III}}^{\pm}(\omega)}{-2\pi} & \sim & \sum_{\gamma=\pm}\int_{0}^{\delta}dk\int_{\gamma k_{0}-\frac{\delta_{z}}{2}}^{\gamma k_{0}+\frac{\delta_{z}}{2}}dk_{z}\frac{k\left(\omega\pm\left(m-B_{z}k_{0}^{2}\right)\right)}{k^{2}+k_{z}^{2}+m_{k}^{2}-\omega^{2}}\nonumber \\
 & \sim & \text{const.}\text{ as }|\omega|\rightarrow\Delta^{-}
\end{eqnarray}
where $\Delta=\frac{1}{2|\bar{B}|}\sqrt{4m\bar{B}-1}$, where $\bar{B}=B(B_{z})$
at region II(III). Here $\delta,\delta_{z}$ are momentum cutoff as
in Eq.\ \eqref{g2d1}. We notice that logarithmic divergence in $g^{\pm}(\omega\rightarrow\pm\Delta)$
is found for TI in region II which is absent in usual semiconductors
which exist only at $m\bar{B}<0$. The sign-inversion effect in $g_{\text{I}}^{\pm}$
is also clear although there is no divergence compared with its 2D
counterpart.
\end{subequations}

\subsubsection{enhanced impurity scattering}

The divergence behavior of $g^{\pm}(\omega)$ has strong implications
on impurity scattering behavior described by the $T$-matrix\ (\ref{t-matrix})
where we observe that impurity scattering can be strongly enhanced
if $g^{\pm}(\omega)$ diverges. In particular TI's with non-diverging
$g^{\pm}(\omega)$'s are more robust towards impurity scattering and
are better candidates for transport application.

Thus (3D) materials belonging to region I and III are the preferred
materials for transport application. For region II, due to the {}``dimension
reduction'' effect, divergence appears near the band edges at both
$\omega\rightarrow\pm\Delta$ and impurity scattering effects are
much enhanced compared with usual semi-conductors. Notice that materials
where all $B$'s are not equal do not suffer from {}``dimension reduction''
effect and are \char`\"{}preferred\char`\"{} also from the point of
view of impurity scattering.

We now apply our analysis to the three TI materials given in Ref.\ \onlinecite{PhysRevB.82.045122},
where the low energy spectra around the inverted bands are given by
\begin{equation}
\omega_{k}=\epsilon_{k}\pm\sqrt{A_{0}^{2}k_{\parallel}^{2}+B_{0}^{2}k_{z}^{2}+M_{k}^{2}}
\end{equation}
 where $\epsilon_{k}=C_{1}k_{z}^{2}+C_{2}k_{\parallel}^{2}$, $M_{k}=M_{0}+M_{1}k_{z}^{2}+M_{2}k_{\parallel}^{2}$
and $k_{\parallel}^{2}=k_{x}^{2}+k_{y}^{2}$. Here we neglect the
higher order term $H_{3}$ which is relevant only for large $k$.
Using the fitting parameters given in Ref.\ \onlinecite{PhysRevB.82.045122}
we classify both the upper and lower topological bands for the three
materials in Table \ref{tab:band-extremum}. We see from Table \ref{tab:band-extremum}
that $\text{Bi}{}_{2}\text{Se}_{3}$ and $\text{Bi}_{2}\text{Te}_{3}$
both have type II band structure and $\text{Sb}_{2}\text{Te}_{3}$
is the preferred material among the three as far as transport application
is concerned.

\begin{table}
\begin{tabular}{|c|c|c|c|c|c|c|}
\hline 
material  & \multicolumn{2}{c|}{$\text{Bi}{}_{2}\text{Se}_{3}$} & \multicolumn{2}{c|}{$\text{Bi}_{2}\text{Te}_{3}$ } & \multicolumn{2}{c|}{$\text{Sb}_{2}\text{Te}_{3}$}\tabularnewline
\hline 
\hline 
band  & upper  & lower  & upper  & lower  & upper  & lower\tabularnewline
\hline 
$k_{0}$  & 0$\AA^{-1}$  & 0.09$\AA^{-1}$  & 0$\AA^{-1}$  & 0.04$\AA^{-1}$  & 0.11$\AA^{-1}$  & 0.08$\AA^{-1}$\tabularnewline
\hline 
Region & I  & II  & I  & II  & III  & III\tabularnewline
\hline 
\end{tabular}

\caption{\label{tab:band-extremum}Band extrema $k_{0}$ and corresponding
types in Fig.\ \ref{fig:band-min} of the inverted band for $\text{Bi}{}_{2}\text{Se}_{3}$,
$\text{Bi}_{2}\text{Te}_{3}$ and $\text{Sb}_{2}\text{Te}_{3}$ (Band
parameters are coming from Ref.\ \onlinecite{PhysRevB.82.045122}).}
\end{table}

\subsection*{Impurity-induced in-gap bound states}

In this section we consider the formation of impurity induced bound
states in the bulk where we shall illustrate the existence of an interesting
band-inversion effect. First we focus on the intra-Dirac bands scattering.
The scattering matrix respecting $\mathcal{T}$ can be written in
orbital basis (the same basis as used in writing down Eq.\ \eqref{eq:G_0})
as 
\begin{equation}
\hat{u}=\left[\begin{array}{cccc}
u_{11} & u_{I} & 0 & \tilde{u}_{I}\\
u_{I}^{*} & u_{22} & -\tilde{u}_{I} & 0\\
0 & -\tilde{u}_{I}^{*} & u_{11} & u_{I}^{*}\\
\tilde{u}_{I}^{*} & 0 & u_{I} & u_{22}
\end{array}\right],\label{eq:u}
\end{equation}
 where $u_{11,22}\in\mathbb{R}$ and $u_{I},\tilde{u}_{I}\in\mathbb{C}$.
To see the qualitative effect of band structure and TI-OI inversion
on bound state formation we consider the case of single scattering
channels, i.e.\ when only one of the $u$'s in Eq.\ \eqref{eq:u}
is non-zero. In this case it is straightforward to obtain the following
eigenvalue equations 
\begin{subequations}
\begin{eqnarray}
\left(1-u_{11}g^{+}(E)\right)^{2} & = & 0\label{eq:u_s}\\
\left(1-u_{22}g^{-}(E)\right)^{2} & = & 0\label{eq:u_p}\\
\left(1-\left|u_{I}\right|^{2}g^{+}(E)g^{-}(E)\right)^{2} & = & 0\\
\left(1-\left|\tilde{u}_{I}\right|^{2}g^{+}(E)g^{-}(E)\right)^{2} & = & 0
\end{eqnarray}
 
\end{subequations}
for each non-zero $u$'s where $E$ is the bound state energy. We
shall analyse in detail the above equations for the case $2mB<1$
in 2D. The analysis can be easily generalized to the case $2mB>1$
and to three dimensions as we shall see in the following.

We start with inter-Dirac-band scattering. Since $g^{+}(E)g^{-}(E)<0$
(c.f. Eq.~\eqref{g2d1}) in the gap region $|E|<|m|$, the scattering
matrix $T$ has no divergence and there is no bound state induced
by inter-Dirac-band scattering. The situation is very different for
intra-band scattering where existence of in-gap bound state depends
on the sign of $u_{ii}g^{\pm}(E)$, and it is easy to see that a bound
state which appears in the conduction/valence band in OI ($m<0$)
state disappears in the corresponding TI ($m>0$) state and vice versa,
where we have again fixed $B>0$ for brevity. The results of bound
state formation is summarized in Table \ref{tab:intra}. The doubly
degeneracy of bound state solutions is a result of Kramer's degeneracy
coming from time reversal symmetry. The existence of bound states
in these cases is a natural result of an attractive (repulsive) impurity
in electron (hole) liquid, which is known to induce bound state in
two dimensions for arbitrarily weak potential. The sign change in
$m$ just reverses the conduction band to valence band or vice versa
but the signs of $u$'s are not reversed, resulting in the appearance/disappearance
of impurity bound states when $m$ changes sign.

\begin{table}
\begin{centering}
\begin{tabular}{|c|c|c|}
\hline 
\multicolumn{1}{|c||}{$u_{11}\neq0$ (Eq.\ \eqref{eq:u_s})} & $u_{11}>0$  & $u_{11}<0$\tabularnewline
\hline 
\hline 
$m>0$  & No  & Yes\tabularnewline
\hline 
$m<0$  & Yes  & No\tabularnewline
\hline 
\end{tabular}
\par\end{centering}

\begin{centering}
\medskip{}

\par\end{centering}

\begin{centering}
\begin{tabular}{|c|c|c|}
\hline 
\multicolumn{1}{|c||}{$u_{22}\neq0$ (Eq.\ \eqref{eq:u_p})} & $u_{22}>0$  & $u_{22}<0$\tabularnewline
\hline 
\hline 
$m>0$  & Yes  & No\tabularnewline
\hline 
$m<0$  & No  & Yes\tabularnewline
\hline 
\end{tabular}
\par\end{centering}

\caption{\label{tab:intra}Single channel intra-Dirac bands scattering induces
in-gap bound state in the weak scattering limit. $m$ is the {}``gap''
parameter in the modified-Dirac Hamiltonian.}
\end{table}

We next consider impurity-induced inter-band scattering between the
Dirac-bands and an extra quadratic band described by Eq.\ (\ref{eq:inter}).
For simplicity we again consider single scattering channel with either
$v_{+(-)}$ being non-zero. It is straightforward to obtain the eigenvalue
equations 
\begin{subequations}
\begin{eqnarray}
\left(1-v_{+}^{2}g^{+}(E)g_{0}^{\text{q}}(E)\right)^{2} & = & 0\label{eq:v_s}\\
\left(1-v_{-}^{2}g^{-}(E)g_{0}^{\text{q}}(E)\right)^{2} & = & 0\label{eq:v_p}
\end{eqnarray}
where we again notice the Kramer's degeneracy of the solutions. Notice
that unlike inter-Dirac-bands scattering where $g^{+}(\omega)g^{-}(\omega)<0$,
independent of the sign of $m$, $g^{\pm}(E)g_{0}^{\text{q}}(E)$
depends now on the sign of $m$. The resulting bound state formation
possibilities are summarized in Table \ref{tab:inter} where we have
considered the quadratic band to be either a conduction or valence
band. Notice that the results are independent of the sign of the impurity
potential since $v_{\sigma}$ always enters the eigenvalue equation
as $v_{\sigma}^{2}$. Again if we fix the scattering channel and the
quadratic band, we find that the bound state appears/disappears when
$m$ changes sign because the sign of $g^{+(-)}(E)$ changes upon
orbital inversion.
\end{subequations}
Our analysis can be extended easily to the case $2mB>1$. There is
no band-inversion effect in this case since ordinary insulator exists
only for $mB<0$ and impurity bound state always exist because of
the {}``double-divergence\char`\"{} behavior in $g^{\pm}(\omega)$.
The situation for 3D TI's are similar. Band inversion effect exist
only in Region I and weak impurity bound state always exist in Region
II.

\begin{table}[H]

\begin{centering}
\begin{tabular}{|c|c|c|}
\hline 
$v_{+}\neq0$ (Eq.\ \eqref{eq:v_s})  & $M,\mu>0$  & $M,\mu<0$\tabularnewline
\hline 
\hline 
$m>0$  & Yes  & No\tabularnewline
\hline 
$m<0$  & No  & Yes\tabularnewline
\hline 
\end{tabular}\medskip{}
\begin{tabular}{|c|c|c|}
\hline 
$v_{-}\neq0$ (Eq.\ \eqref{eq:v_p})  & $M,\mu>0$  & $M,\mu<0$\tabularnewline
\hline 
\hline 
$m>0$  & No  & Yes\tabularnewline
\hline 
$m<0$  & Yes  & No\tabularnewline
\hline 
\end{tabular}
\par\end{centering}

\caption{\label{tab:inter}Induced in-gap bound states resulting from weak
inter-Dirac-quadratic band scatterings. Quadratic bands are described
by $M,\mu>0$ (conduction) or $<0$ (valence), and $m$ is the {}``gap''
parameter of the modified-Dirac equation.}
\end{table}

\subsection{Summary}

We are now in the position to discuss and summarize our results. We
show that impurity scattering effect in TI with band structure parameters
in certain region are enhanced compared with usual semi-conductors.
As a result impurity bound states can form easily in these TI's, leading
to enhanced bulk conductivity and reduced effective band-gap. Impurity
scattering effect can be reduced for TI's with band parameters in
suitable region (region I and III in Fig.\ \ref{fig:band-min} for
3D TI's). Our analysis is applied to three materials $\text{Bi}{}_{2}\text{Se}_{3}$,
$\text{Bi}_{2}\text{Te}_{3}$ and $\text{Sb}_{2}\text{Te}_{3}$ where
we find that $\text{Sb}_{2}\text{Te}_{3}$ is the preferred material
among the three as far as transport application is concerned. We note
that a larger variety of TI materials exist nowadays and our analysis
can be applied if their band structure is known.

We also point out another interesting sign-inversion effect associated
with band inversion in TI-OI transition. The effect is most pronounced
in 2D TI with $2mB<1$. Experimentally, the TI-OI transition can be
controlled by gate voltage in InAs/GaSb quantum well system and the
transport measurement can be carried out at various chemical potentials
inside the band gap\cite{PhysRevLett.107.136603,2011arXiv1106.5819K}.
The band inversion effect can be observed through a change in distribution
of impurity bound state energies when the system changes from TI to
OI state. The effect is strongest if the impurities coupled preferably
to one of the atomic orbitals.

In conclusion, we study the effect of impurity scattering in TI's
in this paper. Our work is complimentary to previous works\cite{PhysRevB.84.035307,2010arXiv1009.5502S,1367-2630-13-10-103016}
that consider impurity bound state forming from continuous deformation
of edge modes. We find that impurity-induced bound state formation
depends strongly on the band structure of the TI's and impurity scattering
can be suppressed in TI's with band structures in the correct region.
Our result provides a guidance for TI material engineering which is
useful for search of applicable TI's.
\begin{acknowledgments}
We acknowledge helpful discussion with S.\ Q.\ Shen. This work is
supported by HKRGC through grant CRF09/HKUST03. 
\end{acknowledgments}
\appendix
\bibliographystyle{apsrev4-1}
\bibliography{ref,ref-TAI,ref-transport}

\begin{thebibliography}{23}%
\makeatletter
\providecommand \@ifxundefined [1]{%
 \@ifx{#1\undefined}
}%
\providecommand \@ifnum [1]{%
 \ifnum #1\expandafter \@firstoftwo
 \else \expandafter \@secondoftwo
 \fi
}%
\providecommand \@ifx [1]{%
 \ifx #1\expandafter \@firstoftwo
 \else \expandafter \@secondoftwo
 \fi
}%
\providecommand \natexlab [1]{#1}%
\providecommand \enquote  [1]{``#1''}%
\providecommand \bibnamefont  [1]{#1}%
\providecommand \bibfnamefont [1]{#1}%
\providecommand \citenamefont [1]{#1}%
\providecommand \href@noop [0]{\@secondoftwo}%
\providecommand \href [0]{\begingroup \@sanitize@url \@href}%
\providecommand \@href[1]{\@@startlink{#1}\@@href}%
\providecommand \@@href[1]{\endgroup#1\@@endlink}%
\providecommand \@sanitize@url [0]{\catcode `\\12\catcode `\$12\catcode
  `\&12\catcode `\#12\catcode `\^12\catcode `\_12\catcode `\%12\relax}%
\providecommand \@@startlink[1]{}%
\providecommand \@@endlink[0]{}%
\providecommand \url  [0]{\begingroup\@sanitize@url \@url }%
\providecommand \@url [1]{\endgroup\@href {#1}{\urlprefix }}%
\providecommand \urlprefix  [0]{URL }%
\providecommand \Eprint [0]{\href }%
\providecommand \doibase [0]{http://dx.doi.org/}%
\providecommand \selectlanguage [0]{\@gobble}%
\providecommand \bibinfo  [0]{\@secondoftwo}%
\providecommand \bibfield  [0]{\@secondoftwo}%
\providecommand \translation [1]{[#1]}%
\providecommand \BibitemOpen [0]{}%
\providecommand \bibitemStop [0]{}%
\providecommand \bibitemNoStop [0]{.\EOS\space}%
\providecommand \EOS [0]{\spacefactor3000\relax}%
\providecommand \BibitemShut  [1]{\csname bibitem#1\endcsname}%
\let\auto@bib@innerbib\@empty
\bibitem [{\citenamefont {Hasan}\ and\ \citenamefont
  {Kane}(2010)}]{RevModPhys.82.3045}%
  \BibitemOpen
  \bibfield  {author} {\bibinfo {author} {\bibfnamefont {M.~Z.}\ \bibnamefont
  {Hasan}}\ and\ \bibinfo {author} {\bibfnamefont {C.~L.}\ \bibnamefont
  {Kane}},\ }\href {\doibase 10.1103/RevModPhys.82.3045} {\bibfield  {journal}
  {\bibinfo  {journal} {Rev. Mod. Phys.}\ }\textbf {\bibinfo {volume} {82}},\
  \bibinfo {pages} {3045} (\bibinfo {year} {2010})}\BibitemShut {NoStop}%
\bibitem [{\citenamefont {Qi}\ and\ \citenamefont
  {Zhang}(2011)}]{RevModPhys.83.1057}%
  \BibitemOpen
  \bibfield  {author} {\bibinfo {author} {\bibfnamefont {X.-L.}\ \bibnamefont
  {Qi}}\ and\ \bibinfo {author} {\bibfnamefont {S.-C.}\ \bibnamefont {Zhang}},\
  }\href {\doibase 10.1103/RevModPhys.83.1057} {\bibfield  {journal} {\bibinfo
  {journal} {Rev. Mod. Phys.}\ }\textbf {\bibinfo {volume} {83}},\ \bibinfo
  {pages} {1057} (\bibinfo {year} {2011})}\BibitemShut {NoStop}%
\bibitem [{\citenamefont {Teo}\ and\ \citenamefont
  {Kane}(2010)}]{PhysRevB.82.115120}%
  \BibitemOpen
  \bibfield  {author} {\bibinfo {author} {\bibfnamefont {J.~C.~Y.}\
  \bibnamefont {Teo}}\ and\ \bibinfo {author} {\bibfnamefont {C.~L.}\
  \bibnamefont {Kane}},\ }\href {\doibase 10.1103/PhysRevB.82.115120}
  {\bibfield  {journal} {\bibinfo  {journal} {Phys. Rev. B}\ }\textbf {\bibinfo
  {volume} {82}},\ \bibinfo {pages} {115120} (\bibinfo {year}
  {2010})}\BibitemShut {NoStop}%
\bibitem [{\citenamefont {Moore}(2010)}]{Moore}%
  \BibitemOpen
  \bibfield  {author} {\bibinfo {author} {\bibfnamefont {J.~E.}\ \bibnamefont
  {Moore}},\ }\href@noop {} {\bibfield  {journal} {\bibinfo  {journal}
  {Nature}\ }\textbf {\bibinfo {volume} {464}},\ \bibinfo {pages} {194}
  (\bibinfo {year} {2010})}\BibitemShut {NoStop}%
\bibitem [{\citenamefont {Checkelsky}\ \emph {et~al.}(2009)\citenamefont
  {Checkelsky}, \citenamefont {Hor}, \citenamefont {Liu}, \citenamefont {Qu},
  \citenamefont {Cava},\ and\ \citenamefont {Ong}}]{Ong-2009}%
  \BibitemOpen
  \bibfield  {author} {\bibinfo {author} {\bibfnamefont {J.~G.}\ \bibnamefont
  {Checkelsky}}, \bibinfo {author} {\bibfnamefont {Y.~S.}\ \bibnamefont {Hor}},
  \bibinfo {author} {\bibfnamefont {M.-H.}\ \bibnamefont {Liu}}, \bibinfo
  {author} {\bibfnamefont {D.-X.}\ \bibnamefont {Qu}}, \bibinfo {author}
  {\bibfnamefont {R.~J.}\ \bibnamefont {Cava}}, \ and\ \bibinfo {author}
  {\bibfnamefont {N.~P.}\ \bibnamefont {Ong}},\ }\href {\doibase
  10.1103/PhysRevLett.103.246601} {\bibfield  {journal} {\bibinfo  {journal}
  {Phys. Rev. Lett.}\ }\textbf {\bibinfo {volume} {103}},\ \bibinfo {pages}
  {246601} (\bibinfo {year} {2009})}\BibitemShut {NoStop}%
\bibitem [{\citenamefont {Qu}\ \emph {et~al.}(2010)\citenamefont {Qu},
  \citenamefont {Hor}, \citenamefont {Xiong}, \citenamefont {Cava},\ and\
  \citenamefont {Ong}}]{Qu13082010}%
  \BibitemOpen
  \bibfield  {author} {\bibinfo {author} {\bibfnamefont {D.-X.}\ \bibnamefont
  {Qu}}, \bibinfo {author} {\bibfnamefont {Y.~S.}\ \bibnamefont {Hor}},
  \bibinfo {author} {\bibfnamefont {J.}~\bibnamefont {Xiong}}, \bibinfo
  {author} {\bibfnamefont {R.~J.}\ \bibnamefont {Cava}}, \ and\ \bibinfo
  {author} {\bibfnamefont {N.~P.}\ \bibnamefont {Ong}},\ }\href {\doibase
  10.1126/science.1189792} {\bibfield  {journal} {\bibinfo  {journal}
  {Science}\ }\textbf {\bibinfo {volume} {329}},\ \bibinfo {pages} {821}
  (\bibinfo {year} {2010})}\BibitemShut {NoStop}%
\bibitem [{\citenamefont {Analytis}\ \emph
  {et~al.}(2010{\natexlab{a}})\citenamefont {Analytis}, \citenamefont
  {McDonald}, \citenamefont {Riggs}, \citenamefont {Chu}, \citenamefont
  {Boebinger},\ and\ \citenamefont {Fisher}}]{Analytis2010}%
  \BibitemOpen
  \bibfield  {author} {\bibinfo {author} {\bibfnamefont {J.~G.}\ \bibnamefont
  {Analytis}}, \bibinfo {author} {\bibfnamefont {R.~D.}\ \bibnamefont
  {McDonald}}, \bibinfo {author} {\bibfnamefont {S.~C.}\ \bibnamefont {Riggs}},
  \bibinfo {author} {\bibfnamefont {J.-H.}\ \bibnamefont {Chu}}, \bibinfo
  {author} {\bibfnamefont {G.~S.}\ \bibnamefont {Boebinger}}, \ and\ \bibinfo
  {author} {\bibfnamefont {I.~R.}\ \bibnamefont {Fisher}},\ }\href
  {http://dx.doi.org/10.1038/nphys1861} {\bibfield  {journal} {\bibinfo
  {journal} {Nat Phys}\ }\textbf {\bibinfo {volume} {6}},\ \bibinfo {pages}
  {960} (\bibinfo {year} {2010}{\natexlab{a}})}\BibitemShut {NoStop}%
\bibitem [{\citenamefont {Analytis}\ \emph
  {et~al.}(2010{\natexlab{b}})\citenamefont {Analytis}, \citenamefont {Chu},
  \citenamefont {Chen}, \citenamefont {Corredor}, \citenamefont {McDonald},
  \citenamefont {Shen},\ and\ \citenamefont {Fisher}}]{Analytis-PRB2010}%
  \BibitemOpen
  \bibfield  {author} {\bibinfo {author} {\bibfnamefont {J.~G.}\ \bibnamefont
  {Analytis}}, \bibinfo {author} {\bibfnamefont {J.-H.}\ \bibnamefont {Chu}},
  \bibinfo {author} {\bibfnamefont {Y.}~\bibnamefont {Chen}}, \bibinfo {author}
  {\bibfnamefont {F.}~\bibnamefont {Corredor}}, \bibinfo {author}
  {\bibfnamefont {R.~D.}\ \bibnamefont {McDonald}}, \bibinfo {author}
  {\bibfnamefont {Z.~X.}\ \bibnamefont {Shen}}, \ and\ \bibinfo {author}
  {\bibfnamefont {I.~R.}\ \bibnamefont {Fisher}},\ }\href {\doibase
  10.1103/PhysRevB.81.205407} {\bibfield  {journal} {\bibinfo  {journal} {Phys.
  Rev. B}\ }\textbf {\bibinfo {volume} {81}},\ \bibinfo {pages} {205407}
  (\bibinfo {year} {2010}{\natexlab{b}})}\BibitemShut {NoStop}%
\bibitem [{\citenamefont {Eto}\ \emph {et~al.}(2010)\citenamefont {Eto},
  \citenamefont {Ren}, \citenamefont {Taskin}, \citenamefont {Segawa},\ and\
  \citenamefont {Ando}}]{Ando-2010}%
  \BibitemOpen
  \bibfield  {author} {\bibinfo {author} {\bibfnamefont {K.}~\bibnamefont
  {Eto}}, \bibinfo {author} {\bibfnamefont {Z.}~\bibnamefont {Ren}}, \bibinfo
  {author} {\bibfnamefont {A.~A.}\ \bibnamefont {Taskin}}, \bibinfo {author}
  {\bibfnamefont {K.}~\bibnamefont {Segawa}}, \ and\ \bibinfo {author}
  {\bibfnamefont {Y.}~\bibnamefont {Ando}},\ }\href {\doibase
  10.1103/PhysRevB.81.195309} {\bibfield  {journal} {\bibinfo  {journal} {Phys.
  Rev. B}\ }\textbf {\bibinfo {volume} {81}},\ \bibinfo {pages} {195309}
  (\bibinfo {year} {2010})}\BibitemShut {NoStop}%
\bibitem [{\citenamefont {Butch}\ \emph {et~al.}(2010)\citenamefont {Butch},
  \citenamefont {Kirshenbaum}, \citenamefont {Syers}, \citenamefont {Sushkov},
  \citenamefont {Jenkins}, \citenamefont {Drew},\ and\ \citenamefont
  {Paglione}}]{Butch-2010}%
  \BibitemOpen
  \bibfield  {author} {\bibinfo {author} {\bibfnamefont {N.~P.}\ \bibnamefont
  {Butch}}, \bibinfo {author} {\bibfnamefont {K.}~\bibnamefont {Kirshenbaum}},
  \bibinfo {author} {\bibfnamefont {P.}~\bibnamefont {Syers}}, \bibinfo
  {author} {\bibfnamefont {A.~B.}\ \bibnamefont {Sushkov}}, \bibinfo {author}
  {\bibfnamefont {G.~S.}\ \bibnamefont {Jenkins}}, \bibinfo {author}
  {\bibfnamefont {H.~D.}\ \bibnamefont {Drew}}, \ and\ \bibinfo {author}
  {\bibfnamefont {J.}~\bibnamefont {Paglione}},\ }\href {\doibase
  10.1103/PhysRevB.81.241301} {\bibfield  {journal} {\bibinfo  {journal} {Phys.
  Rev. B}\ }\textbf {\bibinfo {volume} {81}},\ \bibinfo {pages} {241301}
  (\bibinfo {year} {2010})}\BibitemShut {NoStop}%
\bibitem [{\citenamefont {Ren}\ \emph {et~al.}(2010)\citenamefont {Ren},
  \citenamefont {Taskin}, \citenamefont {Sasaki}, \citenamefont {Segawa},\ and\
  \citenamefont {Ando}}]{PhysRevB.82.241306}%
  \BibitemOpen
  \bibfield  {author} {\bibinfo {author} {\bibfnamefont {Z.}~\bibnamefont
  {Ren}}, \bibinfo {author} {\bibfnamefont {A.~A.}\ \bibnamefont {Taskin}},
  \bibinfo {author} {\bibfnamefont {S.}~\bibnamefont {Sasaki}}, \bibinfo
  {author} {\bibfnamefont {K.}~\bibnamefont {Segawa}}, \ and\ \bibinfo {author}
  {\bibfnamefont {Y.}~\bibnamefont {Ando}},\ }\href {\doibase
  10.1103/PhysRevB.82.241306} {\bibfield  {journal} {\bibinfo  {journal} {Phys.
  Rev. B}\ }\textbf {\bibinfo {volume} {82}},\ \bibinfo {pages} {241306}
  (\bibinfo {year} {2010})}\BibitemShut {NoStop}%
\bibitem [{\citenamefont {Ren}\ \emph {et~al.}(2011)\citenamefont {Ren},
  \citenamefont {Taskin}, \citenamefont {Sasaki}, \citenamefont {Segawa},\ and\
  \citenamefont {Ando}}]{PhysRevB.84.165311}%
  \BibitemOpen
  \bibfield  {author} {\bibinfo {author} {\bibfnamefont {Z.}~\bibnamefont
  {Ren}}, \bibinfo {author} {\bibfnamefont {A.~A.}\ \bibnamefont {Taskin}},
  \bibinfo {author} {\bibfnamefont {S.}~\bibnamefont {Sasaki}}, \bibinfo
  {author} {\bibfnamefont {K.}~\bibnamefont {Segawa}}, \ and\ \bibinfo {author}
  {\bibfnamefont {Y.}~\bibnamefont {Ando}},\ }\href {\doibase
  10.1103/PhysRevB.84.165311} {\bibfield  {journal} {\bibinfo  {journal} {Phys.
  Rev. B}\ }\textbf {\bibinfo {volume} {84}},\ \bibinfo {pages} {165311}
  (\bibinfo {year} {2011})}\BibitemShut {NoStop}%
\bibitem [{\citenamefont {{Xiong}}\ \emph {et~al.}(2011)\citenamefont
  {{Xiong}}, \citenamefont {{Petersen}}, \citenamefont {{Qu}}, \citenamefont
  {{Cava}},\ and\ \citenamefont {{Ong}}}]{Ong-arxiv2011}%
  \BibitemOpen
  \bibfield  {author} {\bibinfo {author} {\bibfnamefont {J.}~\bibnamefont
  {{Xiong}}}, \bibinfo {author} {\bibfnamefont {A.~C.}\ \bibnamefont
  {{Petersen}}}, \bibinfo {author} {\bibfnamefont {D.}~\bibnamefont {{Qu}}},
  \bibinfo {author} {\bibfnamefont {R.~J.}\ \bibnamefont {{Cava}}}, \ and\
  \bibinfo {author} {\bibfnamefont {N.~P.}\ \bibnamefont {{Ong}}},\ }\href@noop
  {} {\bibfield  {journal} {\bibinfo  {journal} {ArXiv e-prints}\ } (\bibinfo
  {year} {2011})},\ \Eprint {http://arxiv.org/abs/1101.1315} {arXiv:1101.1315
  [cond-mat.mes-hall]} \BibitemShut {NoStop}%
\bibitem [{\citenamefont {Shan}\ \emph {et~al.}(2011)\citenamefont {Shan},
  \citenamefont {Lu}, \citenamefont {Lu},\ and\ \citenamefont
  {Shen}}]{PhysRevB.84.035307}%
  \BibitemOpen
  \bibfield  {author} {\bibinfo {author} {\bibfnamefont {W.-Y.}\ \bibnamefont
  {Shan}}, \bibinfo {author} {\bibfnamefont {J.}~\bibnamefont {Lu}}, \bibinfo
  {author} {\bibfnamefont {H.-Z.}\ \bibnamefont {Lu}}, \ and\ \bibinfo {author}
  {\bibfnamefont {S.-Q.}\ \bibnamefont {Shen}},\ }\href {\doibase
  10.1103/PhysRevB.84.035307} {\bibfield  {journal} {\bibinfo  {journal} {Phys.
  Rev. B}\ }\textbf {\bibinfo {volume} {84}},\ \bibinfo {pages} {035307}
  (\bibinfo {year} {2011})}\BibitemShut {NoStop}%
\bibitem [{\citenamefont {{Shen}}\ \emph {et~al.}(2010)\citenamefont {{Shen}},
  \citenamefont {{Shan}},\ and\ \citenamefont {{Lu}}}]{2010arXiv1009.5502S}%
  \BibitemOpen
  \bibfield  {author} {\bibinfo {author} {\bibfnamefont {S.-Q.}\ \bibnamefont
  {{Shen}}}, \bibinfo {author} {\bibfnamefont {W.-Y.}\ \bibnamefont {{Shan}}},
  \ and\ \bibinfo {author} {\bibfnamefont {H.-Z.}\ \bibnamefont {{Lu}}},\
  }\href@noop {} {\bibfield  {journal} {\bibinfo  {journal} {ArXiv e-prints}\ }
  (\bibinfo {year} {2010})},\ \Eprint {http://arxiv.org/abs/1009.5502}
  {arXiv:1009.5502 [cond-mat.mes-hall]} \BibitemShut {NoStop}%
\bibitem [{\citenamefont {Lu}\ \emph {et~al.}(2011)\citenamefont {Lu},
  \citenamefont {Shan}, \citenamefont {Lu},\ and\ \citenamefont
  {Shen}}]{1367-2630-13-10-103016}%
  \BibitemOpen
  \bibfield  {author} {\bibinfo {author} {\bibfnamefont {J.}~\bibnamefont
  {Lu}}, \bibinfo {author} {\bibfnamefont {W.-Y.}\ \bibnamefont {Shan}},
  \bibinfo {author} {\bibfnamefont {H.-Z.}\ \bibnamefont {Lu}}, \ and\ \bibinfo
  {author} {\bibfnamefont {S.-Q.}\ \bibnamefont {Shen}},\ }\href
  {http://stacks.iop.org/1367-2630/13/i=10/a=103016} {\bibfield  {journal}
  {\bibinfo  {journal} {New Journal of Physics}\ }\textbf {\bibinfo {volume}
  {13}},\ \bibinfo {pages} {103016} (\bibinfo {year} {2011})}\BibitemShut
  {NoStop}%
\bibitem [{\citenamefont {Fu}\ and\ \citenamefont
  {Kane}(2007)}]{PhysRevB.76.045302}%
  \BibitemOpen
  \bibfield  {author} {\bibinfo {author} {\bibfnamefont {L.}~\bibnamefont
  {Fu}}\ and\ \bibinfo {author} {\bibfnamefont {C.~L.}\ \bibnamefont {Kane}},\
  }\href {\doibase 10.1103/PhysRevB.76.045302} {\bibfield  {journal} {\bibinfo
  {journal} {Phys. Rev. B}\ }\textbf {\bibinfo {volume} {76}},\ \bibinfo
  {pages} {045302} (\bibinfo {year} {2007})}\BibitemShut {NoStop}%
\bibitem [{\citenamefont {Cheianov}\ and\ \citenamefont
  {Fal'ko}(2006)}]{PhysRevLett.97.226801}%
  \BibitemOpen
  \bibfield  {author} {\bibinfo {author} {\bibfnamefont {V.~V.}\ \bibnamefont
  {Cheianov}}\ and\ \bibinfo {author} {\bibfnamefont {V.~I.}\ \bibnamefont
  {Fal'ko}},\ }\href {\doibase 10.1103/PhysRevLett.97.226801} {\bibfield
  {journal} {\bibinfo  {journal} {Phys. Rev. Lett.}\ }\textbf {\bibinfo
  {volume} {97}},\ \bibinfo {pages} {226801} (\bibinfo {year}
  {2006})}\BibitemShut {NoStop}%
\bibitem [{\citenamefont {Balatsky}\ \emph {et~al.}(2006)\citenamefont
  {Balatsky}, \citenamefont {Vekhter},\ and\ \citenamefont
  {Zhu}}]{RevModPhys.78.373}%
  \BibitemOpen
  \bibfield  {author} {\bibinfo {author} {\bibfnamefont {A.~V.}\ \bibnamefont
  {Balatsky}}, \bibinfo {author} {\bibfnamefont {I.}~\bibnamefont {Vekhter}}, \
  and\ \bibinfo {author} {\bibfnamefont {J.-X.}\ \bibnamefont {Zhu}},\ }\href
  {\doibase 10.1103/RevModPhys.78.373} {\bibfield  {journal} {\bibinfo
  {journal} {Rev. Mod. Phys.}\ }\textbf {\bibinfo {volume} {78}},\ \bibinfo
  {pages} {373} (\bibinfo {year} {2006})}\BibitemShut {NoStop}%
\bibitem [{\citenamefont {Ng}\ and\ \citenamefont
  {Avishai}(2009)}]{PhysRevB.80.104504}%
  \BibitemOpen
  \bibfield  {author} {\bibinfo {author} {\bibfnamefont {T.~K.}\ \bibnamefont
  {Ng}}\ and\ \bibinfo {author} {\bibfnamefont {Y.}~\bibnamefont {Avishai}},\
  }\href {\doibase 10.1103/PhysRevB.80.104504} {\bibfield  {journal} {\bibinfo
  {journal} {Phys. Rev. B}\ }\textbf {\bibinfo {volume} {80}},\ \bibinfo
  {pages} {104504} (\bibinfo {year} {2009})}\BibitemShut {NoStop}%
\bibitem [{\citenamefont {Liu}\ \emph {et~al.}(2010)\citenamefont {Liu},
  \citenamefont {Qi}, \citenamefont {Zhang}, \citenamefont {Dai}, \citenamefont
  {Fang},\ and\ \citenamefont {Zhang}}]{PhysRevB.82.045122}%
  \BibitemOpen
  \bibfield  {author} {\bibinfo {author} {\bibfnamefont {C.-X.}\ \bibnamefont
  {Liu}}, \bibinfo {author} {\bibfnamefont {X.-L.}\ \bibnamefont {Qi}},
  \bibinfo {author} {\bibfnamefont {H.}~\bibnamefont {Zhang}}, \bibinfo
  {author} {\bibfnamefont {X.}~\bibnamefont {Dai}}, \bibinfo {author}
  {\bibfnamefont {Z.}~\bibnamefont {Fang}}, \ and\ \bibinfo {author}
  {\bibfnamefont {S.-C.}\ \bibnamefont {Zhang}},\ }\href {\doibase
  10.1103/PhysRevB.82.045122} {\bibfield  {journal} {\bibinfo  {journal} {Phys.
  Rev. B}\ }\textbf {\bibinfo {volume} {82}},\ \bibinfo {pages} {045122}
  (\bibinfo {year} {2010})}\BibitemShut {NoStop}%
\bibitem [{\citenamefont {Knez}\ \emph {et~al.}(2011)\citenamefont {Knez},
  \citenamefont {Du},\ and\ \citenamefont {Sullivan}}]{PhysRevLett.107.136603}%
  \BibitemOpen
  \bibfield  {author} {\bibinfo {author} {\bibfnamefont {I.}~\bibnamefont
  {Knez}}, \bibinfo {author} {\bibfnamefont {R.-R.}\ \bibnamefont {Du}}, \ and\
  \bibinfo {author} {\bibfnamefont {G.}~\bibnamefont {Sullivan}},\ }\href
  {\doibase 10.1103/PhysRevLett.107.136603} {\bibfield  {journal} {\bibinfo
  {journal} {Phys. Rev. Lett.}\ }\textbf {\bibinfo {volume} {107}},\ \bibinfo
  {pages} {136603} (\bibinfo {year} {2011})}\BibitemShut {NoStop}%
\bibitem [{\citenamefont {{Knez}}\ \emph {et~al.}(2011)\citenamefont {{Knez}},
  \citenamefont {{Du}},\ and\ \citenamefont
  {{Sullivan}}}]{2011arXiv1106.5819K}%
  \BibitemOpen
  \bibfield  {author} {\bibinfo {author} {\bibfnamefont {I.}~\bibnamefont
  {{Knez}}}, \bibinfo {author} {\bibfnamefont {R.-R.}\ \bibnamefont {{Du}}}, \
  and\ \bibinfo {author} {\bibfnamefont {G.}~\bibnamefont {{Sullivan}}},\
  }\href@noop {} {\bibfield  {journal} {\bibinfo  {journal} {ArXiv e-prints}\ }
  (\bibinfo {year} {2011})},\ \Eprint {http://arxiv.org/abs/1106.5819}
  {arXiv:1106.5819 [cond-mat.mes-hall]} \BibitemShut {NoStop}%
\end{thebibliography}%

\end{document}